\documentclass[preprint,preprintnumbers,amsmath,amssymb]{revtex4}

\newcommand{\beq}{\begin{equation}}
\newcommand{\eeq}{\end{equation}}

\def\E{{\cal E}}
\def\a{{\alpha}}
\begin{document}


\title{How the Quantum Universe Became Classical}

\author{J.J.Halliwell}%
\affiliation{Blackett Laboratory \\ Imperial College \\ London SW7
2BZ \\ UK }

\begin{abstract}
This is an informal introduction to the ideas of decoherence and
emergent classicality, including a simple account of the
decoherent histories approach to quantum theory. It is aimed at
undergraduates with a basic appreciation of quantum theory. The
emphasis is on simple physical ideas and pictures.

\end{abstract}

\maketitle




\section*{1. Introduction}

Quantum theory has been around now for almost a century and it is
probably fair to say that, in its most elaborate form, quantum
field theory, it is the most spectacular physical theory every
invented. It applies to and explains essentially all atomic
phenomena and its predictions, some of which are remarkably
precise, have been exceptionally well verified by experiment.
Indeed, there is not one shred of experimental evidence to suggest
that the basic structure of quantum theory is in any way
incorrect. Suitably adapted, it is expected to apply to all types
of matter and all types of forces holding matter together, from
the very smallest scales of fundamental particles, right up to the
cosmological scale, and possibly even to the entire universe. Any
theory claiming to be fundamental must be expressed in the
language of quantum theory. It is {\it the} fundamental theory.

Yet, despite being the fundamental theory of matter, quantum
theory is not the theory of the ordinary everyday world of our
immediate experience. Many of its features, such as superposition
states and entanglement, defy intuition in a truly profound way,
to the extent that that are even described as ``paradoxical".
Furthermore, the mathematical objects of its vocabulary do not
correspond directly to the things we see and measure in the world
about us, and the question of the relationship between the
formalism and physical reality -- the interpretation of the theory
-- has been a hotly debated topic.

These tensions are due to the fact that the large scale world of
immediate experience is best described by other, older areas of
physics, namely the classical mechanics of Newton, devised in the
18th century, and the theory of thermodynamics, devised in the
19th century. These classical theories are theories of the
ordinary everyday world, and the mathematical objects they deal
with do correspond directly to things we see and measure. They are
utterly different in their structure to quantum theory.

This then leads to a very interesting question. If large scale
objects are made of atoms, and atoms are described by quantum
theory, how do large collections of atoms come to be described by
classical physics? In brief, how does classical physics {\it
emerge} from quantum theory?

This question has been one of great interest to many researchers
over the last ten or twenty years, and the purpose of this paper
is to explain what this question means and how one can begin to
solve it. The emphasis will be on simple physical ideas and
pictures, not on elaborate mathematics, in the hope that it will
understandable to the broadest possible audience.

\section*{2. Classical Physics}

A very wide variety of particle systems at ordinary scales are
very accurately described by the classical mechanics of Newton.
For, say, a single particle system, the description is very
simple. The state of the system at each time is characterized by
its position $x$ and momentum $p$ of each particle.  If the forces
are specified through a potential $V (x)$, then the evolution in
time is described by Newton's law
\begin{equation}
m { d^2 x \over dt^2} + V'(x) =0
\label{newton}
\end{equation}
We may solve this equations, on a computer if necessary, and the
solution gives a curve from which we may see the position and
momentum at any time, given the initial position and momentum. The
laws describing this very simple system readily generalize to
systems consisting of many particles with all sorts of complicated
interactions between them. They describe a wide variety of
phenomena at small scales almost as small as the atomic scale,
right up to the scale of planets, stars and the whole universe.

The mathematical objects the Newtonian theory deals with
correspond directly with our immediate experience of the world:
position, momentum, acceleration, force and so on. One can think
of a classical system as having ``definite properties'', such as definite
position and momentum, without contradiction. There is no sense in
which the theory needs to be ``interpreted''.

Classical mechanics also readily generalizes to stochastic
theories. These are theories where one has incomplete information
about some of the variables and they have to be modelled in terms
of random processes. A simple example is Brownian motion, where a
small particle, like a particle of pollen, is seen to jiggle
around as a result of invisible collisions with molecules. The
pollen particle approximately follows the equations of Newtonian
mechanics, but with random fluctuations around that motion and
also with dissipation. In one dimension, this is described by an
equation of the form,
\begin{equation}
m { d^2 x \over dt^2} + m \gamma { dx \over d t} + V'(x) =  noise
\label{stoch}
\end{equation}
Other important examples, where the fluctuations are more
significant, are the stock market or the weather. In these
stochastic theories we still think of the variables like position
and momentum as taking definite values, even though those values
are known imperfectly.

\section*{3. Quantum Theory}

Classical physics was realized to be inadequate around the
beginning of the 20th century when physicists started trying to
understand the structure of atoms. In particular, it did not
explain the striking fact that radiation emitted from atoms occurs
only at discrete sets of frequencies. Niels Bohr in 1913 put
forward a model of the atom in which the electrons orbited the
nucleus at circular orbits of fixed size and the discrete nature
of the radiated spectrum was explained by the transition between
these orbits. Eventually, however, the founders of quantum theory
made a much bolder and more far-reaching assumption, namely that
elementary particles, like electrons, are best modelled by the
mathematics of waves. This explained the discrete set of sizes for
the orbits since a whole number of waves had to fit around the
atom, and moreover, the possible frequencies of vibration of these
waves explained the spectrum of radiation. Hence the discreteness
comes from the natural relationship between waves and discrete
sets of numbers.

Absolutely central to this description is an equation determining
the precise form of these waves. This is Schr\"odinger's famous
equation,
\begin{equation}
i \hbar \frac { \partial \psi} {\partial t } = \left( - \frac
{\hbar^2 } {2m} \nabla^2 + V({\bf x}) \right) \psi
\end{equation}
perhaps the single most important equation in the whole of
physics. It makes an utterly spectacular range of predictions and
there is absolutely no doubt that it is ``correct'' in any sense
of the word.

The solution to the Schr\"odinger equation is the wave function $\psi ({\bf x},t)$, a
complex-valued field which describes the possible frequencies of
vibration of the atoms and the precise way in which they vibrate.
However, although the vibrational modes appeared to be
naturally related to the frequencies of atomic radiation, the
introduction of the wave function as a description of atomic
matter was completely new. This quantity does not relate directly
to measurements, or to our intuitive understanding of the physical
world. It requires an {\it interpretation}.

An interpretation of a theory becomes necessary when the
mathematical objects in its vocabulary do not directly correspond
to the things we measure or to our intuitive picture of the
physical world. Classical mechanics does not need an
interpretation, since the objects in its basic vocabulary --
position, momentum, energy, force -- correspond to our direct
experience of the physical world and to the things we measure. In
fact, the whole idea of interpretation is essentially meaningless
in classical mechanics. However, quantum mechanics introduces a
genuinely new concept, the wave function, which does not
immediately correspond to anything we can measure or have a
physical feeling for. In some ways, it is remarkable that quantum
theory was discovered at all!

What do we mean by ``interpretation''? This is a deep question
which has some difficult philosophical aspects, but broadly, it
comes down to two things. Firstly, it refers to the practical way
in which physical predictions corresponding to measurements are
extracted from the wave function. Secondly, and more
philosophically, the phrase ``interpretation'' refers also to the
picture of reality suggested by the theory. Quantum theory appears
to involve a very strange picture of reality at the atomic level
and this is one reason it is so difficult to really understand.
Here, we will for the large part be concentrating on the first
more pragmatic question, although it is probably reasonable to say
that views on the second issue underlie all thinking in this area,
if only unconsciously.

\section*{4. The Probability Formula}

If the electrons around an atom are described in terms of a wave
that is spread out all over the place, it became important to
understand what this means, given that previously an electron was
thought of as a point particle at a specific point in space. It
was quickly realized in the development of quantum theory
that the wave function $\psi (x)$ may be
interpreted in terms of probability. Particles no longer could be
thought of as being at definite points in space -- there is only a
probability of finding it at a certain point. Furthermore,
the probability is given in terms of the wave function by the
important formula
\begin{equation}
| \psi (x) |^2 = {\rm  probability \ of \ finding \ the \ particle
\ at \ the \ point} \ x
\label{born}
\end{equation}
However, the founders of quantum mechanics did not hit on this
straight away. In the original paper of Max Born in 1926 \cite{Born}, who
first put forward this idea, we find the statement,
that the probability is proportional to a quantity he denotes
$ \Phi_{n,m} $ (essentially the same as $ | \psi |$ in the notation used here).
But then
there is a footnote, clearly added at a later stage just before publication,
stating, ``more careful consideration shows that the probability
is proportional to the square of the quantity $\Phi_{n,m}$''.
Born still got the Nobel Prize anyway, despite this minor hitch!

Max Born may well have been guessing when he put forward
this interpretation, but now, in the modern understanding of
quantum theory, the theorem of Gleason confirms that the Born
interpretation is essentially the only reasonable one \cite{Gle}. The fact
that the probabilities are given by the square of the wave function
leads to much of the mystery and apparent
paradox of quantum theory, that is, to the phenomena that are
utterly different to classical phenomena. More about this shortly.

The wave description of matter and its probability interpretation
automatically leads to the Heisenberg uncertainty principle. This
says that the position $x$ and momentum $p$ of a particle cannot
be precisely specified but suffer uncertainties $ \Delta p$ and
$\Delta x $ which obey the inequality
\begin{equation}
(\Delta p) (\Delta x) \ge {\hbar \over 2}
\label{UP}
\end{equation}
where $\hbar$ is Planck's constant and is of order $10^{-34} Js$.
Planck's constant is clearly a very small number compared to
ordinary laboratory scales, which shows why this effect is not
seen in classical physics. The uncertainty principle alone
indicates that the description of physics systems suffers from
statistical uncertainties, similar to those encountered in
stochastic classical theories. However, what is important is that
quantum mechanics is not just a classical stochastic theory, but
exhibits new effects which are provably distinct from merely
stochastic effects.

\section*{5. Superposition States and the Double-Slit
Experiment}

What makes quantum theory truly different from classical physics
is the possibility of superposition states. The Schr\"odinger
equation is a linear wave equation. This means that if $\psi_1$
and $\psi_2 $ are solutions to it, describing two different
physical situations, then the superposition state
\begin{equation}
\psi = \psi_1 + \psi_2
\label{super}
\end{equation}
is also a solution to the equation -- it is also a valid physical
state. An example of such a superposition state is shown in Figure
1.

So far, there is nothing remarkable here, since most wave
equations (like those obeyed by sound waves) have the same
property. However, in quantum mechanics, the physics
depends on the probabilities, which are given by the {\it square}
of the wave function, $ | \psi |^2$, and this does not behave a
simple way for superposition states. For the superposition state
Eq.(\ref{super}), we have
\begin{equation}
| \psi |^2 = | \psi_1^2 | + | \psi_2 |^2 + \psi_1 \psi_2^* +
\psi_1^* \psi_2
\end{equation}
which means that the probability $ | \psi |^2 $ associated with
the total wave function is not simply related to the probabilities
$ | \psi_1 |^2 $ and $ | \psi_2 |^2 $ of the individual elements of
the superposition. This leads to an important observation: that
one of the essential differences between classical and quantum
theory is that in the quantum theory, probabilities do not add up
in the way we expect them to on the basis of classical intuition.

We can state this even more sharply. Mathematically, there is a
close relationship between probability theory and logic. Loosely
speaking, when probabilities can be assigned and add up properly,
then we can make logical statements about the system that may be
manipulated according to the rules of ordinary logic. This is
sometimes called Boolean logic, or classical logic -- basically,
it is the logic of ordinary language. We probabilities cannot
be assigned then the logical rules of ordinary language are not
necessarily respected. So quantum theory has the possibility
of violating the familiar logical structure of ordinary
language.
Hence, another way of expressing the essential difference between
classical and quantum mechanics is to say that they involve
different types of logic, and that quantum mechanics does not
respect the logic of ordinary everyday language. The phrase
``quantum logic'' is sometimes used to describe the logical rules
respected by quantum theory, but this does not help us to
understand it.

An outstandingly important example of these ideas is the double
slit experiment with electrons, first performed by Davisson and
Germer in 1927 \cite{DaG}. In this experiment, electrons are fired
from a source towards a blind with two holes in it, and then
detected at a screen beyond the holes, as shown in Figure 2. What
the detector in fact measures is an interference pattern, and this
experiment was a major piece of evidence in favour of the wavelike
nature of matter.

The wave function $ \psi (x)$ for particles that arrive at the
screen at point $x$ is given by a linear superposition of wave
functions coming from hole $1$ and hole $2$. The probability of
arriving at a certain point on the screen is given by the square
of this,
\begin{equation}
P_{12} = | \psi_1 + \psi_2 |^2
\label{pattern}
\end{equation}
and this indeed is what is measured by the detector.

If we were to apply our classical intuition to this experiment, we
would be inclined to say that, since the particle hit the screen
at $x$ and since it was at an earlier time emitted from the
source, then it must have passed through either slit number $1$ or
slit number $2$. If this were the case, then we could assign
probabilities $P_1 = | \psi_1 |^2$ and $P_2 = | \psi_2|^2$ to the
electrons coming from each slit, and the probability of hitting
the screen would be simply
\begin{equation}
P_1 + P_2 = | \psi_1 |^2 + | \psi_2 |^2
\end{equation}
But this is not the same as Eq.(\ref{pattern}), the result
that is actual measured! So
the probabilities don't add up properly and we quite simply cannot
make the logical deduction that the electron went through one hole
or the other. The rules of ordinary logic do not apply! This
simple experiment and its interpretation is perhaps the biggest
challenge quantum theory makes on our intuition.
(Interestingly, in
a recent poll in Physics World, readers voted the double slit
experiment with single electrons as ``the most beautiful
experiment in physics'' \cite{IOP}.)

Other interesting challenges to our intuition become possible
when we consider wave functions describing two particles.
In particular, states of the form
\begin{equation}
\Psi (x_1, x_2) = \phi (x_1) \chi (x_2) + \phi' (x_1) \chi' (x_2)
\label{ent}
\end{equation}
are called {\it entangled} states and can often involve a degree
of correlation between the particles which cannot be understood
in classical terms. The EPR state is the most famous
example of such a state \cite{EPR}.

\section*{6. The Copenhagen Interpretation}

We now need to say a little bit more about the interpretation of
quantum theory. The probability formula is not the end of the
story, but it is part of a wider interpretational framework put
forward by Bohr in the 1920s, which became known as the Copenhagen
interpretation. This set of ideas, which is widely accepted for
most practical purposes, describes how quantum systems are
affected by measurements \cite{WhZ}.

In the strongest statement of the Copenhagen interpretation, it is
asserted that the world is divided into {\it two} different
regimes. There is the atomic scale, described by quantum
mechanics, and the macroscropic, laboratory scale, which, to agree
with our experience, must be described by the classical mechanics
of Newton. In some sense we never in fact apprehend the quantum
world directly, only its indirect effects on the macroscopic,
classical world. The existence of a classical regime was thought
to be necessary for the interpretation of the theory. This is
already
rather unsatisfactory since we would like quantum theory to
be fundamental, yet we seem to have to postulate a division of the world
into classical and quantum realms in order to understand quantum theory.
It gets even more complicated.

The Copenhagen interpretation gives a specific mechanism for the
way in which quantum systems are affected as a result of being
measured.
Suppose, for example, the quantum state of the system is a
superposition state in which a single particle is localized
about two different positions, as in Figure 1. If we perform a measurement
which asks whether the particle is in a certain region,
$\Delta$, say, then, according to the Copenhagen interpretation,
the effect of the measurement is that the wave function outside that region
gets squashed to zero, leaving behind only the part within
$\Delta$. See Figure 3.
This is the famous ``collapse of the wave function''
process. It is instantaneous and discontinuous.
Furthemore, the precise way in which the wave function
changes depends on {\it what} is measured --
it changes differently, for
example, if one measures momentum or energy instead of position.

Quantum systems therefore have two modes of evolution: smooth
evolution according to the Schr\"odinger equation interspersed with
abrupt jumps when measurements take place.
Hence the evolution of quantum systems depends on whether they
are being observed and what is being observed.
David
Mermin once wrote an article entitled, ``Is the Moon Really There
When No-One Looks?'', elaborating on this disturbing aspect
of quantum systems \cite{Mer}.
The apparent observer-dependence of quantum
theory is yet another fact that is utterly different to classical
mechanics, since there is no inconsistency in maintaining that
the classical
systems of our immediate experience continue to exist whether or not we
look at them, and that looking at them does not change them.


\section*{7. How is Classical Mechanics Different to Quantum
Mechanics?}

We now then come to the turning point of this article, which is to
emphasize the difference between classical and quantum theory.
These differences are best summarized in the following table.

\bigskip
\bigskip

\begin{tabular}{|l|l|}
\hline
CLASSICAL MECHANICS & QUANTUM MECHANICS \\
\hline
system state: $p,x$ & system state: $ \psi (x)$ \\
\hline
$p,x$ take definite values & superpositions, entanglement
\\ & $(\Delta
p) (\Delta x) \ge {\hbar \over  2} $ \\
\hline
classical (Boolean) logic & quantum logic\\
\hline
$ m \ddot x + V'(x) = 0$ & $i \hbar {\partial \psi \over \partial t}
= H \psi $ \\
\hline
observer-independent & depends on act of measurement \\
\hline
no need for interpretation & Copenhagen interpretation \\
\hline
\end{tabular}

\bigskip
\bigskip
We have arrived at the following picture: Classical mechanics has a
broad but finite degree of applicability. Quantum theory
supersedes it as the description of nature in a much wider class
of situations. Its description of the world is extremely different
to that provided by classical mechanics. But they both describe
the same phenomena. How can one reduce to the other? Furthermore,
quantum mechanics needs a classical domain to assist its
interpretation, yet this leads to a rather unsatisfactory
dualistic view of the world.

It seems that the way out of this is to unashamedly assume that
quantum theory is the fundamental description of everything, from
the tiniest atom right up to the entire universe \cite{Everett}. This explicitly
drops the assumption of a separate external classical domain
together with the rather cumbersome machinery describing quantum measurements in
the Copenhagen interpretation. Our
task, then, is to show that substantial parts of the universe
appear to be classical even though they are described
fundamentally by quantum mechanics.

This now brings us to the main point of this article, in which we consider
how quantum theory becomes approximated by classical theory under
certain conditions.

\section*{8. Emergent Classicality: Some Clues}

In the standard quantum mechanics text books, it is often
asserted that classical mechanics ``emerges''
from quantum mechanics when we take the quantum mechanical
averages of position and momentum, $ \langle x \rangle $, $\langle
p \rangle $. Indeed, it follows from the Schrodinger equation that
$ \langle x \rangle $ obeys the equation of motion
\begin{equation}
m { d^2   \langle x \rangle \over dt^2 }  + \langle V'(x) \rangle
= 0
\end{equation}
which is clearly very similar to the classical equation of
motion Eq.(\ref{newton}) for $\langle x \rangle$. This is the famous Ehrenfest
theorem. But it is not exactly the same unless,
\begin{equation}
\langle V'(x) \rangle = V' ( \langle x \rangle )
\label{av}
\end{equation}
which is only true exactly for linear systems.

Another relevant fact is the evolution
of wavepackets strongly concentrated about particular values
of position and momentum (but consistent with the uncertainty principle).
Such wavepackets satisfy Eq.(\ref{av}) approximately, and under evolution
according to the Scr\"odinger equation, follow approximately classical trajectories,
at least for a period of time. However, these wavepackets
typically spread out in time, which means that their positions
and momentum become somewhat blurry unlike macroscopic classical
objects. See Figure 4.

Despite their limitations, these results do however provide some useful
clues. They show that there is an approximate classical determinism
buried in the Schr\"odinger equation and that this might emerge
if we look at the system in a sufficiently crude or averaged
sort of way, that is, if we take a {\it coarse-grained} view of the dynamics.
In simple terms, this is like looking at something from so far away that it is
impossible to see the finer details of the structure. The necessity for coarse-graining
in order to obtain classical physics is to be expected anyway because of the uncertainty
principle, Eq.(\ref{UP}), since it is only by looking at position and momentum
very coarsely that the effects of the uncertainty principle are not noticed.

However, the really important difference between classical and quantum systems is that
quantum systems may exist in superposition states, exhibiting interference
effects, and the Ehrenfest theorem says nothing about how these disappear.

\section*{9. Interferences and Classical States}

The most important issue in understanding the appearance of classical
mechanics from quantum mechanics is seeing how interferences are suppressed.
This issue is most usefully rephrased in terms of the density matrix
(and now we need to briefly get a bit more mathematical).
For a system described in terms of a wave function $\psi (x)$, we define the
density matrix by
\begin{equation}
\rho (x,y) = \psi (x) \psi^* (y)
\label{pure}
\end{equation}
So far there is nothing new in this.
However, the density matrix is most useful when both quantum and statistical
effects are separately present in the system. Suppose, for example, we don't
know exactly which quantum state the system is in, but we know only the
probability $p_n$ that the system is in the quantum state $ \psi_n (x)$.
Then a very useful object is the density matrix
\begin{equation}
\rho (x,y) = \sum_n p_n \psi_n (x) \psi^*_n (y)
\end{equation}
The quantum description of thermal equilibrium, for example, involves
a state of this form.
A state of this type, in which they are a combination of
quantum and statistical effects is called a mixed state, whereas Eq.(\ref{pure}),
which contains only quantum effects is called a pure state.

Now, the density matrix also gives a particularly useful picture of superposition
states. Suppose the system is in a superposition state,
\begin{equation}
\psi (x) = { 1 \over \sqrt{2} } \left( \psi_1 (x) + \psi_2 (x) \right)
\end{equation}
(where the $  { 1 \over \sqrt{2} } $ factor is for normalization).
Then the associated density matrix is
\begin{equation}
\rho(x,y) = \frac {1} {2} \left[ \psi_1 (x) \psi_1^* (y) + \psi_2 (x) \psi_2^* (y) + \psi_1
(x) \psi_2^* (y) + \psi_2 (x) \psi_1^* (y) \right]
\label{rho1}
\end{equation}
The last two terms represent the interferences, and because of them,
we are not allowed to say that the system is in state $\psi_1 $ or $\psi_2$
(because remember that the probabilities don't add up properly). To get emergent
classicality, what we want is for these terms to go away, and to get instead the mixed
state,
\begin{equation}
\rho(x,y) = \frac {1} {2} \left[ \psi_1 (x) \psi_1^* (y) + \psi_2
(x) \psi_2^* (y) \right]
\label{rho2}
\end{equation}
This density matrix {\it does} correspond to the statement that the system is in state
$\psi_1$ or state $\psi_2$, with probability $1/2$ for being in either state.

For the superposition state shown in Figure 1, a superposition of spatially
localized states, the two density matrices Eqs.(\ref{rho1}), (\ref{rho2}) are
shown in Figures 5 and 6.

\section*{10. Measurements, Decoherence and the Environment}

The key question now is, how do the interference terms
in the density matrix go away?
The first question one might ask is whether the interference terms
naturally go away under evolution according to the Schr\"odinger
equation. There are a variety of ways of showing that this is in
fact impossible. It is quite easy to show that the property of
purity is preserved by the Schr\"odinger equation, so a pure state
cannot go to a mixed state. Physically, this is because a mixed
state involves statistical effects, whereas a pure state does
not. To go from a pure state to a mixed state, statistical
effects need to appear somehow, but the Schr\"odinger equation
contains no such effects. (Although it is interesting to note
that explicit modifications of the Scrh\"odinger equation, which
contain precisely such effects have been postulated \cite{GRW}).
Something more ingenious is therefore required.

At this stage, standard measurement
theory in the Copenhagen approach gives a key insight. Recall that in the Copenhagen
approach, it does not seem to be necessary to worry about
the non-classical effects of
superpositions. We just {\it measure} the system, whatever the state
is, and get the result, which is then regarded as a ``classical'' thing.
That is, when a quantum system is measured, in the Copenhagen
sense, parts of superposition states go away. See Figure 3. But here, recall,
we are
trying to avoid talking about measurements because we are taking
quantum theory to be universal. This brings us to an important
idea. The standard machinery of quantum mechanics with the Schr\"odinger
equation applies to isolated systems, but in practice, true
isolation is extremely hard to achieve for macroscopic systems. The
vast majority of macroscopic physical systems are not in fact isolated but are
in continual interaction with their immediate surroundings -- their
environment, as we call it. The environment in effect continually
``measures''  macroscopic systems and it seems reasonable to suppose
that it is this effect that habitually destroys interference,
thereby rendering the systems classical. This effect is called {\it
decoherence} and is absolutely central to our understanding of
emergent classicality. See Figure 7.

\section*{11. System-Environment Models}

These compelling physical ideas about decoherence are nicely illustrated in the
classic calculation of Joos and Zeh \cite{JoZ} who considered the evolution
of a massive particle, such as a dust grain, being continuously ``measured'' by its
interactions with the surrounding gas molecules. They considered a
particle with position $x$ in interaction with an environment $
\E$. The wave function for the combined system and environment is
$ \psi (x, \E)$ and this wave function describes the situation in
which the dust grain is correlated with the surrounding
molecules causing the measurements of it.

However, the key idea here is that the details of the
environment state are not in fact of interest, so we average
over it (an example of coarse-graining).
The state of the system only is then described by an object called the
reduced density matrix, obtained by averaging over environment
states:
\begin{equation}
\rho (x,y) = \sum_{\E } \psi (x, \E ) \psi^* (y,\E)
\end{equation}
This object is similar in interpretation to the usual density
matrix
in standard quantum theory. However, because we have averaged out
part of the system it no longer evolves according to the
Schr\"odinger equation. The whole system described by $\psi (x, \E)$
evolves according to the Schr\"odinger equation, but for the
particle only, Joos and Zeh derived a so-called master equation,
whose one-dimensional version is
\begin{equation}
{ \partial \rho \over \partial t}  = - {i \hbar \over 2 m } \left(
{\partial^2 \rho \over \partial x^2} -  { \partial^2 \rho \over
\partial y^2} \right) - D (x-y)^2 \rho
\label{master}
\end{equation}
The first term on the right-hand side is easily recognized as the
usual Schr\"odinger evolution. What is new is the second term, $ D
(x-y)^2 \rho (x,y)$, describing the continuous measurement effect of the
environment. Importantly, this term turns pure states into mixed
states, exactly what we need.

How does this new term affect the dynamics? On short timescales we
have
\begin{equation}
\rho (x,y,t) \ \sim \ e^{ - D (x-y)^2 t } \rho (x,y,0)
\label{diag}
\end{equation}
which means that the interference terms with $x \ne y $ decay
exponentially fast, the effect we are looking for. In
typical models,
$$
D = { 2 m \gamma k T \over \hbar^2}
$$
where $T$ is the temperature of the environment and $\gamma$ the
dissipation. The key thing is that $ D $ goes like $ 1 / \hbar^2$
so $D$ has the possibility of being extremely large for laboratory scales
and hence the interference terms extremely small. For example,
if we take the mass, temperature, lengthscale $ | x-y|$, etc. to be of order
$1$ in cgs units, we find that the off-diagonal terms, representing
interference, are very small indeed:
\begin{equation}
e^{ - D (x-y)^2 t } \  \sim  \ \ \exp
\left(-10^{40} \right) \ \
\end{equation}
This shows rather decisively why interference effects are not observed
on macroscopics scales! See Ref.\cite{Zur} for a similar calculation.
Note also that the decoherence effect may also be shown to kill
the non-classical effects of entangled states, such as Eq.(\ref{ent})
\cite{Dio,HaD}.

This example is by no means pathological.
For most macroscopic systems decoherence is the most ruthless and efficient
process in the whole of physics and is the main physical explanation as
to how quantum systems become classical.

\section*{11. What does it mean for a quantum system to be
approximately classical?}

We now have a heuristic picture of how we can understand the
emergence of classical mechanics from quantum theory. It is
time to focus this into something more precise.

When we say that a system is classical, we mean that it can be
described by variables $p,x$ which have definite values (or at
least, probabilities) and that these variables evolve according to
classical evolution equations. Classical evolution equations, such
as Newton's law, Eq.(\ref{newton}),
involve {\it two time derivatives}. So to assert that the classical
equation of motion is satisfied we need to specify position at
three moments of time. This means that the whole notion of ``classical''
is tied up with the notion of a history -- a sequence of events
distributed in time.

This suggests that if we wish to decide whether a given quantum
system is approximately classical, we do the following.
We use quantum theory to attempt calculate the probability for a
history of positions, $p (\alpha_1, t_1, \alpha_2, t_2 \cdots )$
and then see if it is strongly peaked about the classical
evolution equations. See Figure 8. However, it is here that the interesting
tension lies, because quantum theory resists the notion of
history. This is for two reasons.

First, the uncertainty principle limits the precision to
within which positions at different times may be specified (since
this is equivalent to specifying position and momentum at one
time). This means that probabilities for histories cannot be
perfectly peaked about a single classical path.
However, if we look at positions that are
sufficiently coarse-grained compared to the uncertainty principle
limits, we might expect the probability to be quite strongly peaked
about one path.

Secondly, and more fundamentally, as we have seen in the double
slit experiment interference resists the assignment of
probabilities to histories. However, we know we can get round this
when there is a decoherence mechanism to kill the interferences.

It therefore appears to be a very worthwhile goal explore
formulations of quantum mechanics which concentrates on looking at
probabilities for histories, and which also incorporates a
decoherence mechanism to make sure that those probabilities are
well-defined.

\section*{12. The Decoherent Histories Approach to Quantum
Theory}

Everything said so far is very concisely and
comprehensively synthesized in a new understanding of quantum
theory which goes by the name of the decoherent histories approach
(or consistent histories approach). This approach was invented by
Robert Griffiths in 1984 \cite{Gri,Gri2} and subsequently developed
substantially by Roland Omn\`es \cite{Omn,Omn2}. Murray Gell-Mann and James Hartle
invented and developed a similar approach, in part independently \cite{GeH,GeH2}.
(See also Ref.\cite{Hal1} for a recent review).

The decoherent histories approach is standard quantum mechanics
but without the usual assumptions about classical domains or
measurements. It assumes that everything is quantum, up to and
including the entire universe, so there is no division of
the world into classical and quantum regimes.
Its just all quantum theory. Instead of measurements, the theory
concentrates on finding those situations to which probabilities
may be assigned. This is a much weaker and more objective notion
and does not rely on a classical domain. We look at histories
partly because they are the most general class of possible situations,
but also because they are necessary to characterize classical
behaviour we seek to explain the emergence of.

Here is a simple account of how it works. We need to extend the
formalism of quantum theory so that it can describe a history  --
a series of events distributed in time. Suppose the system has an
initial state $ | \psi \rangle $ and we want to talk, in quantum
mechanics, about a history in which the system evolves through a
series of properties $\a_1, \a_2 \cdots \a_n$ at times $t_1, t_2
\cdots t_n$. Then this history is represented by the state
\begin{equation}
|\psi_{\a} \rangle = P_{\a_n}(t_n) P_{\a_{n-1}} (t_{n-1} ) \cdots
P_{\a_1} (t_1) | \psi \rangle
\end{equation}
Here the objects $P_{\a} (t)$ are projection operators at each
moment of time characterizing
the property we are interested in at that time. This is best
thought of in terms of a diagram. Without going into detail, the
state $ |\psi_{\a} \rangle $ could represent, for example, the
statement in quantum mechanics that the particle started out in
the state $\psi (x)$ and then it passed through the gate $\a_1$ at
time $t_1$, and then the gate $\a_2$ at $t_2$ and so on. See Figure 9.
Following
the normal rules of quantum theory, the probability for the entire
history is
\begin{equation}
p(\a_1, \a_2, \cdots \a_n ) = \langle \psi_{\a} |
\psi_{\a} \rangle
\end{equation} which is nothing more than an extension to
histories of the usual Born rule, Eq.(\ref{born}).

However, as already stressed, probabilities cannot in
general be assigned to histories because of interference.
Interference means that probabilities will not add properly in the
way we expect classical probabilities to do, so the probabilities
are meaningless. Now here is the important point. What is new
about the decoherent histories approach is that the theory also
includes a test to see which sets of histories suffer quantum
interference and which do not. The mathematical condition
for no interference is
that all distinct pairs of histories $\a, \a'$ in the set must
have no overlap:
\begin{equation}
\langle \psi_{\a'} | \psi_{\a} \rangle = 0 \ \ \ \ {\rm for} \ \ \
\a \ne \a'
\label{deco}
\end{equation}
The more physical significance of this condition is that it ensures
that the probabilities
add up properly, that is, that
\begin{equation}
p( A  \ or \ B ) = p (A) + p(B)
\label{prob}
\end{equation}
for all exclusive pairs of distinct histories $A$ and $B$.
When histories satisfy this condition, we call them a decoherent
set of histories. They can be assigned probabilities and
manipulated as it they were classical histories. There is no need
to appeal to notions of measurement.

We know from the double slit experiment that that no-interference condition
Eq.(\ref{deco}) or Eq.(\ref{prob}) will not be satisfied in general, so
probabilities cannot always be assigned to histories. But it can be
satisfied when there is an environment present to produce decoherence,
since satisfaction of Eq.(\ref{deco}) is very similar to the diagonalization
of the density matrix Eq.(\ref{diag}). When probabilities can be
assigned we can talk about the properties of the system as if they
were real, classical things. Differently put, in the non-Boolean
logic of quantum theory, we look for Boolean subsets.

So this is the histories approach in a nutshell: a formula for
probabilities for histories and a simple condition to determine
when those probabilities are meaningful.

This basic framework has been used extensively to analyze emergent
classicality in a variety of different settings \cite{GeH2,Hal1}.
For example,
one can compute the probability that a particle passes through
a series of spatial regions at a sequence of time, with an
environment present to ensure decoherence. The probabilities
for such histories are typically strongly peaked about classical
equations of motion, not quite of the form Eq.(\ref{newton}),
but with dissipation. Furthermore, as discussed above, the peaking
cannot be perfect so there are fluctuations about the classical
path (produced by the essentially random effects of the environment).
Hence what emerges from quantum theory is not precisely
Newtonian mechanics, Eq.(\ref{newton}), but a classical stochastic
theory, like Eq.(\ref{stoch}). For particles of sufficiently
large mass, the stochastic effects are however negligible and
a near-deterministic equation of motion is obtained.

The decoherent histories approach, however, offers much more than
a coherent account of emergent classicality. Since the formalism does not
rely in measurements it can be used to analyze conceptually tricky
points in quantum theory where the debate centres around what is
happening to the system when it is not being measured. Of course,
the decoherent histories approach cannot provide answers where
none exist any more than standard quantum theory can. However, the
emphasis on finding situations where Boolean logic applies brings
a certain clarity lacking in the Copenhagen interpretation. Often
so-called ``paradoxes'' in quantum theory are seen to arise from a
use of logic that, in the decoherent histories approach, would not
be permitted \cite{Gri,Gri2,Omn,Omn2}.


\section*{13. The Preferred Basis Problem}

There is another important issue in the question of emergent
classicality still to be addressed. This is the issue of
why certain variables, such as position, momentum and energy
appear to have a preferred role in describing the classical world,
when quantum theory appears to indicate no preference. That is, in
the decoherence process that produces classical behaviour,
interferences between different values of {\it position} are
killed, and so we get to assign probabilities to histories of
position. Quantum theory doesn't care if the state of the system
is localized in position or in a superposition of such localized
states but somehow localized position states appear to be
preferred.

When decoherence is produced by interactions with an environment,
the provisional answer to this question is that the environment
interacts with the system of interest through position. The
interactions are {\it local} in position. So the environment
effectively ``measures'' the position of the system and not some other
funny variable.

But there is a lot more to it than that. The question relates to
an even more general question, which is how do we decide where to
draw the line between ``system'' and ``environment''? In many
situations there is no obvious environment to produce the
decoherence process. This takes us right back to where we started,
to the old Copenhagen interpretation, in which the world was
arbitrarily divided into classical and quantum regimes.

Fortunately there is the beginnings of an answer to these issues,
and it appears most clearly in the decoherent histories approach.
The decoherent histories approach to quantum theory does not
specifically require an environment to produce decoherence,
just some sort of coarse-graining process.
There is in fact a set of general physical principles that help
us to decide which variables
to concentrate on and which to ignore in the coarse graining
process necessary to get decoherence, namely, {\it conservation laws}.
For a system
consisting of a collection of interacting particles, the total
energy, momentum, number, charge etc. are conserved, that is, they
are unchanged under time evolution. Importantly, conservation laws
persist through to quantum theory (usually), and at least in
simple situations in non-relativistic physics, energy, momentum,
number etc. are conserved.

Now in quantum theory, conserved quantities have a very
interesting property. This is that there can be no interference
between different values of a conserved quantity. The reason for
this is that to observe the interference effects between, say,
states of different charge, one would need a measuring device
which violates charge conservation, and this is impossible.
Alternatively, if we have a superposition state of conserved
quantities (as in Figure 10, for example), the different
elements of the superposition
can interfere only if they evolve into each other. But
conserved quantities don't evolve anywhere, even in the quantum theory
so the different elements of the superposition never meet.
So the superposition still ``exists'' in some sense, but its
interference effects are never seen.


In the decoherent histories approach, this means that one can {\it
always} assign probabilities to histories of conserved quantities.
Or in other words, conserved quantities can always be manipulated
as if they were classical, even in the quantum theory. Now this is
a very important point. It means that in the foggy landscape of
quantum theory, the conserved quantities represent the mountain
peaks that reach above the clouds, immune to weird quantum
behaviour.

Conserved quantities are not the end of the story. They can be
used to define quantities that are not exactly conserved, but
approximately conserved, and so slowly varying. In particular, we
can define the local densities associated with them: if energy,
for example, is conserved, then the energy in a small region of
the system can only change by flowing in or our of the region, so
will tend to change slowly. See Figure 11. If exactly conserved quantities are
decoherent then approximately conserved quantities
will be approximately decoherent \cite{GeH2,hydro}. This means that, to a very good
approximation, we can assign probabilities to them, which will add
up properly with only very small errors.

Hence the real question is not how to divide up the universe
into classical and quantum, or system and environment, but into
slowly and rapidly varying, and conservation laws are the
fundamental physical principle that guides us in how to do this.

These ideas have been used to give what is perhaps the most
general possible account of emergent classicality from quantum
theory. That is, the approximate conservation of the local
densities has been used to show that they are naturally
decoherent, and to obtain classical equations of motion for them.
These
equations are hydrodynamics equations, of the Navier-Stokes form,
for example \cite{hydro,hydro2}.

\section*{14. Summary}

The essence of this article may be very concisely summarized as follows:

\noindent{\bf (a)} Classical physics emerges from quantum theory at sufficiently coarse-grained scales
when there is a natural division into fast and slow variables.

\noindent{\bf (b)} Conservation laws are the guiding principle in making that division.

\section*{15. Other Literature}

The point of this article has been to give a simple flavour of the
physical ideas involved in understanding emergent classicality, mainly from
the perspective of the decoherent histories approach. It
is far from exhaustive and the selection of references is rather
sparse. Much work has been done on many issues touched on here.
See the review article Ref.\cite{Hal1} for a more extensive review
of the decoherent histories approach, with references. Many of the key
ideas about emergent classicality are nicely covered in the article
by Hartle. See also the websites maintained by Brun
\cite{Bru} and by Joos \cite{Joos}, both very valuable and extensive
sources of literature.

\section*{Acknowledgements}

I have benefited from discussions with many people over the years on the topic
of this paper, but special mention goes to Jim Hartle for many conversations
over a very long period of time. I would also like to thank
Neal Powell and Marko Ivin for their assistance
in preparing the figures.

\section*{Figure Captions}

\noindent{\bf Figure 1.} A wave function consisting of a superposition of two wave
functions, each localized around a different point in space.

\noindent{\bf Figure 2.} The double slit experiment. Electrons are
fired at a blind with two holes in it and detected at the screen.
At the screen, the wave function of the system is of the form $
\psi = \psi_1 + \psi_2 $, where $ \psi_1$ and $\psi_2$ represent
the wave functions of electrons coming from slits $1$ and $2$
respectively. The detector measures an intensity proportional to $
P_{12} = | \psi_1 + \psi_2 |^2 $. This is is not the same as
$ | \psi_1 |^2 + | \psi_2 |^2$, the result one might guess from
classical expectations.

\noindent{\bf Figure 3.} A schematic representation of the collapse of the wave function
process which takes place as a result of a quantum measurement. A measurement localized
around the right-hand peak in the superposition state causes the wave function to
``collapse'' to zero everywhere outside the measured region, leaving only the
piece of the wave function within the measured region.

\noindent{\bf Figure 4.} The spreading of the wave function. A wavepacket
tightly peaked in position at $t=1$ evolves along an approximately
classical trajectory but tends to become more spread out as time evolves.

\noindent{\bf Figure 5.} The density matrix for a pure state, Eq.(\ref{rho1}),
for the superposition state of the
form shown in Figure 1. The interference terms are represented by the two
off-diagonal $( x \ne y) $ peaks.

\noindent{\bf Figure 6.} The density matrix for the mixed state,
Eq.(\ref{rho2}). The off-diagonal terms representing interferences
are absent compared to Figure 5.

\noindent{\bf Figure 7.} An example of a system being continually monitored
by its environment. The system is a collection of small particles, such
as dust grains, and the environment consists of the molecules in the surrounding
atmosphere which scatter off the dust grains, effectively ``measuring'' them.

\noindent{\bf Figure 8.} The trajectory of a classical system is determined
by seeing if it passes through a series
of gates $\a_1, \a_2, \a_3 \cdots $ at times $ t_1, t_2, t_3...$.

\noindent{\bf Figure 9.} A schematic picture of the construction of the
quantum state for a history $ | \psi_{\a} \rangle $. It represents
the statement that the system starts in the initial
wave function $ | \psi \rangle $ and then
passes through the gates $ \a_1, \a_2, \a_3, \cdots $ at times
$t_1, t_2, t_3 \cdots $.

\noindent{\bf Figure 10.} A wave function consisting of a superposition
of two states, each strongly concentrated around different values
of energy.

\noindent{\bf Figure 11.} A volume $V$ of a box of gas molecules.
Since energy (for example) is conserved, the total energy within
$V$ can only change by particles flowing in or our of $V$. If $V$ is
sufficiently large compared to the microscopic scale (and comoving
with the average flow), then the total energy in $V$ will be a slowly
varying quantity, so will we approximately decoherent.

\section*{About the Author}

J.J.Halliwell did his PhD research with Stephen Hawking
at the University of Cambridge, graduating in 1986. After
a Research Fellowship for a year at Christ's College, Cambridge,
he moved to the USA, spending two years at the Institute
for Theoretical Physics, Santa Barbara, and three years at
MIT, in Cambridge, Massachusetts. He came to Imperial College
in 1992, initially as a Royal Society University Research Fellow,
and shortly after was appointed lecturer. He has been a Professor
of Theoretical Physics since 2002. His research interests are centred
around quantum cosmology, quantum gravity and the related problems
these fields generate in the foundations of quantum theory.

\end{document}